\documentclass[prd,nofootinbib,superscriptaddress,twocolumn]{revtex4}

\usepackage{graphicx}
\usepackage[breaklinks=true,colorlinks=true,linkcolor=blue,urlcolor=blue,
citecolor=blue]{hyperref}

\usepackage[mathscr]{eucal}
\usepackage[latin1]{inputenc}
\usepackage{amsmath}
\usepackage{amsfonts}
\usepackage{amssymb}
\usepackage{pictex}
\usepackage{listings}
\usepackage{tcolorbox}

\def\db{\delta_b}
\def\dc{\delta_c}

\def\be{\nopagebreak[3]\begin{equation}}
\def\ee{\end{equation}}
\def\ba{\nopagebreak[3]\begin{eqnarray}}
\def\ea{\end{eqnarray}}

\newcommand{\teta}{\rlap{\lower2ex\hbox{$\,\tilde{}$}}\eta{}}

\def\lp{{\ell}_{\rm Pl}}

\usepackage{enumerate}

\begin{document}

\title{Basis Function Method for Numerical Loop Quantum Cosmology: \\ The Schwarzschild Black Hole Interior}
\author{Alec Yonika}
\affiliation{Department of Physics \& Center for Scientific Computing and Visualization Research,
      University of Massachusetts Dartmouth, North Dartmouth, MA 02747, USA}
\author{Gaurav Khanna}
\affiliation{Department of Physics \& Center for Scientific Computing and Visualization Research,
      University of Massachusetts Dartmouth, North Dartmouth, MA 02747, USA}

\begin{abstract}
Loop quantum cosmology is a symmetry-reduced application of loop quantum gravity that has led to the resolution 
of classical singularities such as the big bang, and those at the center of black holes. This can be seen 
through numerical simulations involving the quantum Hamiltonian constraint that is a partial {\em difference} equation. 
The equation allows one to study the evolution of sharply-peaked Gaussian wave packets that generically exhibit a quantum 
``bounce'' or a non-singular passage through the classical singularity, thus offering complete singularity resolution. 
In addition, von-Neumann stability analysis of the difference equation -- treated as a stencil for a numerical solution that 
steps through the triad variables -- yields useful constraints on the model and the allowed space of states. In this paper, 
we develop a new method for the numerical solution of loop quantum cosmology models using a set of basis functions that offer 
a number of advantages over computing a solution by stepping through the triad variables. We use the Corichi and Singh model 
for the Schwarzschild interior as the main case study in this effort. The main advantage of this new method is computational 
efficiency and the ease of parallelization. In addition, we also discuss how the stability analysis appears in the context 
of this new approach. 
\end{abstract}

\maketitle

\section{Introduction}
Classical general relativity is plagued with singularities such as those at the center of black holes and also the big bang in 
cosmological models. In loop quantum gravity, the spacetime continuum is replaced by a discrete quantum structure wherein geometric 
operators such as areas and volumes have discrete eigenvalues with a non-zero minimum~\cite{lqg1,lqg2,lqg3}. The discrete structure of 
quantum spacetime only plays a significant role when the curvature approaches Planck scale; otherwise, strong agreement with 
general relativity is found. Nearly two decades of study has been performed in the context of singularity resolution in 
symmetry-reduced cosmological models with fairly robust results that replace the big bang with a big bounce~\cite{lqc,lqc1,lqc2,lqc3}. 
States sharply peaked on classical trajectories when the universe is large and expanding, can be evolved backward using the quantum 
Hamiltonian constraint; such states evolve in a stable and non-singular way, and bounce in the deep Planck regime into a contracting 
branch~\cite{dgms14}.

Several of the above mentioned results were obtained through the application of computational techniques in the field of loop quantum 
gravity (See Refs.~\cite{nlqc,nlqc1,nlqc2} and references therein). 
Recently, similar numerical studies were performed in the context of the Corichi and Singh (CS) model of the Schwarzschild interior~\cite{CS} 
resulting in a clear numerical demonstration of singularity resolution~\cite{YKS}. The CS model improves upon previous models, offering a 
consistent and correct infra-red limit and independence from fiducial structures used in the quantization procedure. The quantum Hamiltonian 
constraint is a partial {\em difference} equation in two discrete triad variables making it somewhat more complicated in comparison with the 
previously studied cosmological models. Von-Neumann stability analysis of this equation results in a stability condition for black holes 
which have a very large mass compared to the Planck mass. In addition, further analysis for such large black holes leads to a constraint 
on the choice of the allowed states in numerical evolution. Evolution of a sharply peaked Gaussian wave packet yields a bounce in one of the 
triad variables, but for the other triad variable singularity resolution arises through a simple passage through the classical singularity. 
In addition, states are found to be peaked at the classical trajectory for a long time before and after the classical singularity~\cite{YKS}. 
These results support a {\em symmetric} quantum black hole to white hole transition paradigm that has received tremendous attention in recent years~\cite{bhwh,bhwh1,bhwh2}.

Needless to say that the semi-classical behavior of the model is expected in the regime where the triad variables have large values in comparison 
to the Planck scale. This implies that the numerical triad grid required must span a very large domain, and is often limited by the finite 
computational resources (memory, compute time, numerical precision, etc.). In practice, our previous efforts succeeded in performing 
numerical simulations only on the scale of a few hundred Planck units in each triad dimension~\cite{YKS}. The method used therein was a 
straightforward approach inspired by a finite-difference stencil computation -- a recursive stepping through a 2D grid built using two discrete-
valued triad variables. The method was difficult to parallelize owing to its intrinsically serial structure and was also constrained 
by finite floating-point numerical precision. In this paper, we develop a new numerical solution method inspired by the well-known 
spectral collocation approach for numerical solutions of partial differential equations that utilizes a set of basis functions in one triad 
variable, and performs an explicit stepping in the other variable. This allows for high computational efficiency, especially for the larger 
sized computations and reduced numerical precision requirements. Moreover, the basis function approach is readily parallelizable on multi- and 
many- core processors like modern CPUs and GPUs. 

An alternative method of solution to quantum Hamiltonian constraints could also offer a different perspective on the role of the von-Neumann 
stability analysis in loop quantum cosmology models. After all, one may ask, are the discovered instabilities through that analysis a true feature 
of the model and possibly the underlying physics, or simply an artifact of some sort of how one solves the equations? 
The von-Neumann stability analysis is most commonly used to evaluate finite-difference stencil-based computations, i.e. a local stepping approach 
on a grid built using the triad variables' discrete set of values. What if one took a more global approach towards solving the equation, that 
doesn't involve any stepping? Does the same instability manifest itself in some other way? If so, then indeed, that would be a strong indication of 
the inherent nature of the instability and its relevance to a property of the model and possibly even something physical. In this paper, we study this 
question in some detail and show that the previously discovered instabilities are not simply artifacts of the manner in which the equations where 
solved, rather they are indicative of the issues within the CS model itself or something else of significance (see Ref.~\cite{YKS} for different 
possibilities).

This paper is organized as follows: 
In Sec.~\ref{Background} we introduce the loop quantum CS model and briefly describe its key features. In Sec.~\ref{BFM} we present the new 
numerical method to solve such models, and in Sec.~\ref{VS} we apply this technique to a variable-separated representation of the CS model. 
In Sec.~\ref{2DBFM} the application is broadened to the full 2D form of the CS model. A discussion of some of the technical aspects of the 
implementation of the new numerical technique are also presented therein. A detailed discussion of the instability exhibited by the CS model 
is presented in Sec.~\ref{INST}. We conclude with a discussion of results in Sec.~\ref{DnC}, and the novel benefit that the numerical technique 
offers is elaborated on in Sec.~\ref{PARA} and Sec.~\ref{PREC}.

\section{Background}\label{Background}
The loop quantization of the Schwarzschild interior is performed using a Kantowski-Sachs vacuum spacetime with a phase space expressed 
in terms of holonomies of Ashtekar-Barbero connection components $b$ and $c$, and the two conjugate triad variables $p_b$ and $p_c$. The 
space-time metric in terms of these variables is given by
\begin{equation}
 {\mathrm{d}} s^2 = - N^2 {\mathrm{d}} t^2 + \frac{p_b^2}{|p_c| L_o^2} {\mathrm{d}} x^2 + |p_c| ({\mathrm{d}} \theta^2 + \sin^2 \theta {\mathrm{d}} \phi^2) ~.
\end{equation}
Here $L_o$ is a fiducial length scale in the $x$-direction of the spatial manifold. To relate this with the usual Schwarzschild metric 
variables, $p_b$ and $p_c$ satisfy 
\begin{equation}
 \frac{p_b^2}{p_c} = \frac{2 m}{t} - 1 ~~~~ |p_c| = t^2 ~,
\end{equation}
where $m = G M$, with $M$ as the ADM mass of the black hole space-time. In the classical theory, the horizon at $t=2m$ is identified with 
$p_b = 0$ and $p_c = 4 m^2$ while the central singularity is where both $p_b$ and $p_c$ vanish. In the quantum theory, the eigenvalues of triad 
operators are given by
\begin{equation}\label{pb_pc_ev}
\hat p_b \, |{\mu,\tau}\rangle = \frac{\gamma \lp^2}{2} \, \mu \, |{\mu,\tau}\rangle, ~~ \hat p_c \, |{\mu,\tau}\rangle = \gamma \lp^2 \, \tau \, |{\mu,\tau}\rangle 
\end{equation}
where $\gamma$ is the Immirzi parameter and $\lp$ is the Planck length. Loop quantization of the classical Hamiltonian constraint using the holonomies of 
the connection components $b$ and $c$ yields the following quantum difference equation~\cite{CS}
\begin{eqnarray}\label{CS}
&& \nonumber \db(\sqrt{|\tau|} + \sqrt{|\tau + 2 \dc|}) \left(\Psi_{\mu + 2 \db, \tau + 2 \dc} - 
\Psi_{\mu - 2 \db, \tau + 2 \dc} \right) \\
&& \nonumber \hskip-0.2cm + \tfrac{1}{2} \, (\sqrt{|\tau + \dc|} - \sqrt{|\tau -  \dc|}) \bigg[(\mu+2\db)\Psi_{\mu + 4\db, \tau} \\
&& \nonumber \hskip0.4cm ~~~~~ +  (\mu-2\db) 
\Psi_{\mu - 4\db, \tau}  - 2 \mu (1 + 2 \gamma^2 \db^2)  \Psi_{\mu,\tau} \bigg]\\
&& \nonumber \hskip-0.2cm + \db(\sqrt{|\tau|} + \sqrt{|\tau - 2 \dc|}) \left(\Psi_{\mu - 2\db, \tau - 2 \dc} - 
\Psi_{\mu+ 2\db, \tau - 2 \dc} \right) \\
&& = 0 ~.
\end{eqnarray}
Here, for a Schwarzschild black hole interior corresponding to mass $m$,  
\begin{equation}\label{dbdc}
\delta_b = \frac{\sqrt{\Delta}}{2 m}, ~~~ \mathrm{and} ~~~~ \delta_c = \frac{\sqrt{\Delta}}{{L_o}} ~~~ 
\end{equation}
where $\Delta$ denotes the minimum area eigenvalue in loop quantum gravity, i.e. $\Delta = 4 \sqrt{3} \pi \gamma \lp^2$. 

\section{Basis Function Method}\label{BFM}
In this section, we demonstrate the use of our newly developed basis function method (BFM) to solve discrete quantum Hamiltonian constraints in 
loop quantum cosmology. As mentioned earlier, this approach is inspired by the well-known spectral collocation method use commonly for 
numerical solutions of partial differential equations. The main advantage of the basis function method is its high computational efficiency and 
ease of parallelization on modern computer hardware. We also compare the basis method based solutions with the previous approach of recursively  
stepping (RSM) through a grid of triad variable values. 

We use the example of the CS model of the Schwarschild interior throughout. We set $\gamma\db \rightarrow 0$ (a requirement for 
stable solutions~\cite{YKS}) and set $\delta_c = 2\delta_b=1$ without loss of generality. 

\subsection{Separable Solutions}\label{VS}
Let us begin with performing a separation-of-variables solution of the CS model under consideration. To demonstrate the viability of the basis 
function method solution for this model, $\Psi$ is taken to be of the form $\Psi_{\mu,\tau} \rightarrow A(\mu)B(\tau)$, which reduces the CS model to 
\begin{eqnarray}
(\mu + 2\db)A(\mu+4\db) & + & (\mu-2\db)A(\mu-4\db)= 2\mu A(\mu)\nonumber \\ 
2\db\lambda(A(\mu-2\db)& - &A(\mu+2\db)) + 4\mu\gamma^2\db^2 A(\mu)
\label{Aeqn}
\end{eqnarray}
and
\begin{eqnarray}
&(\sqrt{|\tau|}+\sqrt{|\tau + 2\dc|})B(\tau+2\dc)& \nonumber \\ &-(\sqrt{|\tau|}+\sqrt{|\tau-2\dc|})B(\tau-2\dc)& \nonumber \\
&=-\lambda(\sqrt{|\tau+\dc|} - \sqrt{|\tau-\dc|})B(\tau)&
\label{Beqn}
\end{eqnarray}
where $\lambda$ is the separation parameter. 

The basic idea behind the basis function method is to start with a set of appropriate functions, and then solve the difference equation on a linear span of this set. The chosen basis should allow for highly variable behavior for small values of the triad variables $\mu$ and $\tau$, but only capture very smooth behavior for much larger values. This allows for very sharp quantum fluctuations in the deep Planck regime, and yet very smooth semi-classical behavior for large values of the triad variables. One approach to build such a basis is to seek inspiration from a Fourier basis as used in spectral collocation methods, but with a non-constant wavelength, i.e. $k(x)$. Note that our requirement of smoother behavior for larger values of $x$, then becomes the condition that $k(x)$ be a monotonically decreasing function of $x$. Specifically, we let $k$ drop rapidly (exponentially) with $x$. Thus our desired modified Fourier basis would take the form, $\exp(i n \theta(x))$ where $\theta'(x)=k(x)$ drops exponentially with $x$. Given such a form for a basis, we can then expand any solution as a linear combination of these basis elements and obtain the values of the coefficients by imposing the difference equation. Thus, ultimately we solve a linear system of equations for which a wide variety of efficient numerical algorithms and solvers are  readily available.

\begin{center}
\begin{figure}[htb!]
  \includegraphics[width=\columnwidth]{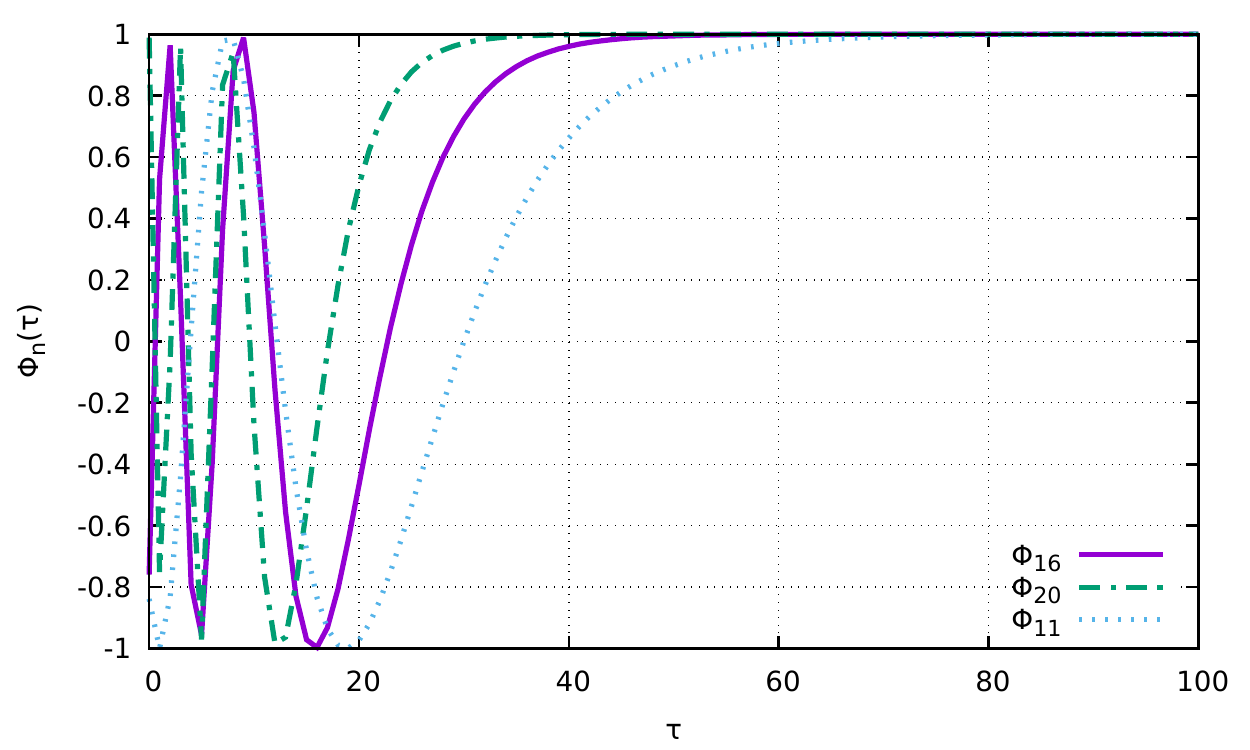}
  \caption{Sample basis elements utilized throught this work. Note how they allow for highly oscillatory behavior at small values of the triad and then smoothen out as the values get larger. }
  \label{fig:basis}
\end{figure}
\end{center}

More specifically, we choose the basis functions to take the form,
\begin{equation}\label{basis}
    \Phi_n(\tau) = \exp(i n e^{-\frac{|\tau|}{N}\exp(\frac{2n}{N})})
\end{equation}
which is a slightly modified version of a basis that was proposed by one of us in Ref.~\cite{CK06} many years ago. Some sample basis elements are depicted in Fig.~\ref{fig:basis}.

The basis function method ultimately involves solving a linear system of equations for a vector of weights $\omega_n$, where $n$ is an index spanning the number of basis elements. The system matrix is represented as an ``interpolation matrix'', with each entry being the sequence in Eqn.~\ref{Beqn}, and with each $B$ replaced with an appropriate representation of Eqn.~\ref{basis}. Each row in this matrix corresponds to an increasing value of $\tau$ in the domain of the computation. Thus, Eqn.~\ref{Beqn} would be represented as 
\begin{eqnarray}
&\sum^N_{n=0} \omega_n\big[(\sqrt{|\tau|}+\sqrt{|\tau + 2\dc|})\Phi_n(\tau+2\dc)& \nonumber \\
&-(\sqrt{|\tau|}+\sqrt{|\tau - 2\dc|})\Phi_n(\tau-2\dc)&\nonumber\\
&+\lambda(\sqrt{|\tau+\dc|} - \sqrt{|\tau-\dc|})\Phi_n(\tau)\big] = 0 & \label{Bseqinterp}
\end{eqnarray}

To solve the above system for the function $B(\tau)$, other than the trivial $B(\tau)=0$, inspiration was taken from a standard technique for numerically solving partial differential equations. A row can be injected into the system matrix, as long as there is an accompanying column, to disrupt the homogeneity of the system without fundamentally changing the computation. This row can be interpreted as numerically incorporating initial or boundary conditions; and in our approach is treated as a restriction on the solution. For our results in this section, we used $\sum_n\omega_n\Phi_n(\tau=T)=1$ for a large value of $T$. 

\begin{center}
\begin{figure}[htb!]
  \includegraphics[width=\columnwidth]{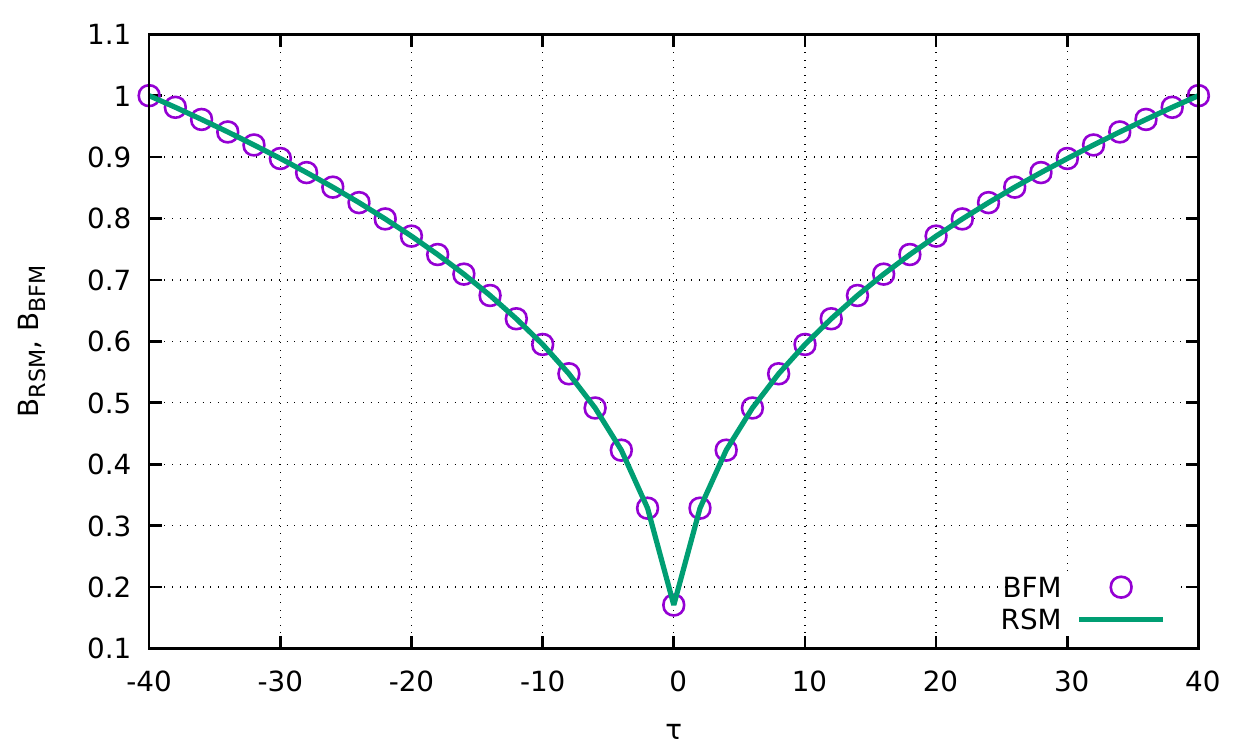}
  \includegraphics[width=\columnwidth]{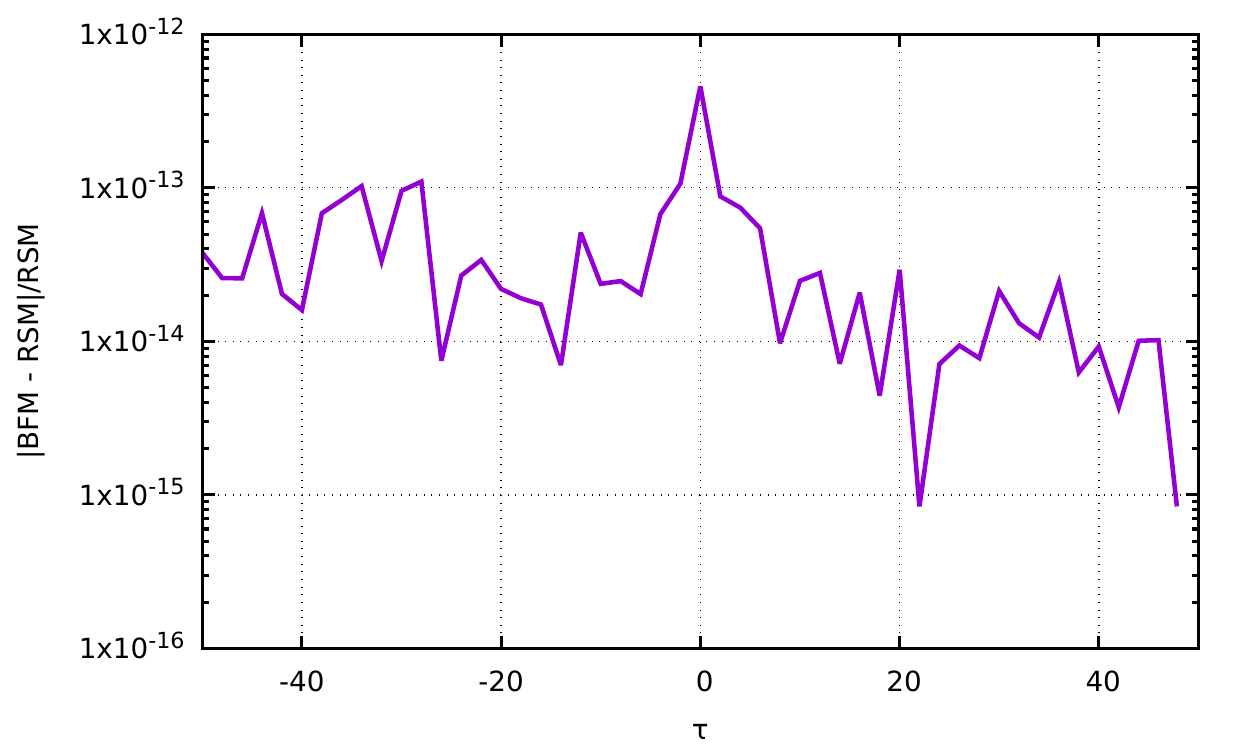}
  \caption{Solution of the $B(\tau)$ equation using the basis function method as compared with a solution computed by stepping through the $\tau$ values using a recursive approach. The upper panel shows both solutions plotted together. The lower panel depicts the absolute difference between the two solutions, which is on the scale of machine precision.}
  \label{fig:fd}
\end{figure}
\end{center}

Using this approach, a reconstructed solution can be generated, which is in precise agreement with the recursively computed solution. In Fig.~\ref{fig:fd} we show a sample basis solution for the $B(\tau)$ equation as compared with a recursion based solution computed by stepping through the values of the $\tau$ variable. There, we set $\gamma\db \rightarrow 0$ (a requirement for stable solutions) and set $\delta_c = 2\delta_b=\lambda=1$ for simplicity.

Similarly to solve Eqn.~\ref{Aeqn} we first make the substitution $C(\mu) \equiv A(\mu+2\db)-A(\mu-2\db)$ which results in 
\begin{eqnarray}
    \mu\big(C(\mu+2\db)-C(\mu-2\db)\big)&+&\nonumber \\
    2\db\big(C(\mu+2\db)+C(\mu-2\db)\big)&=&-2\lambda C(\mu) ~.
\end{eqnarray}
This allows us to easily solve Eqn.~\ref{Aeqn} using both the basis function method and the recursive approach. The outcome is similar as that shown 
previously for the $B(\tau)$ function. 
\begin{center}
\begin{figure}[htb!]
  \includegraphics[width=\columnwidth]{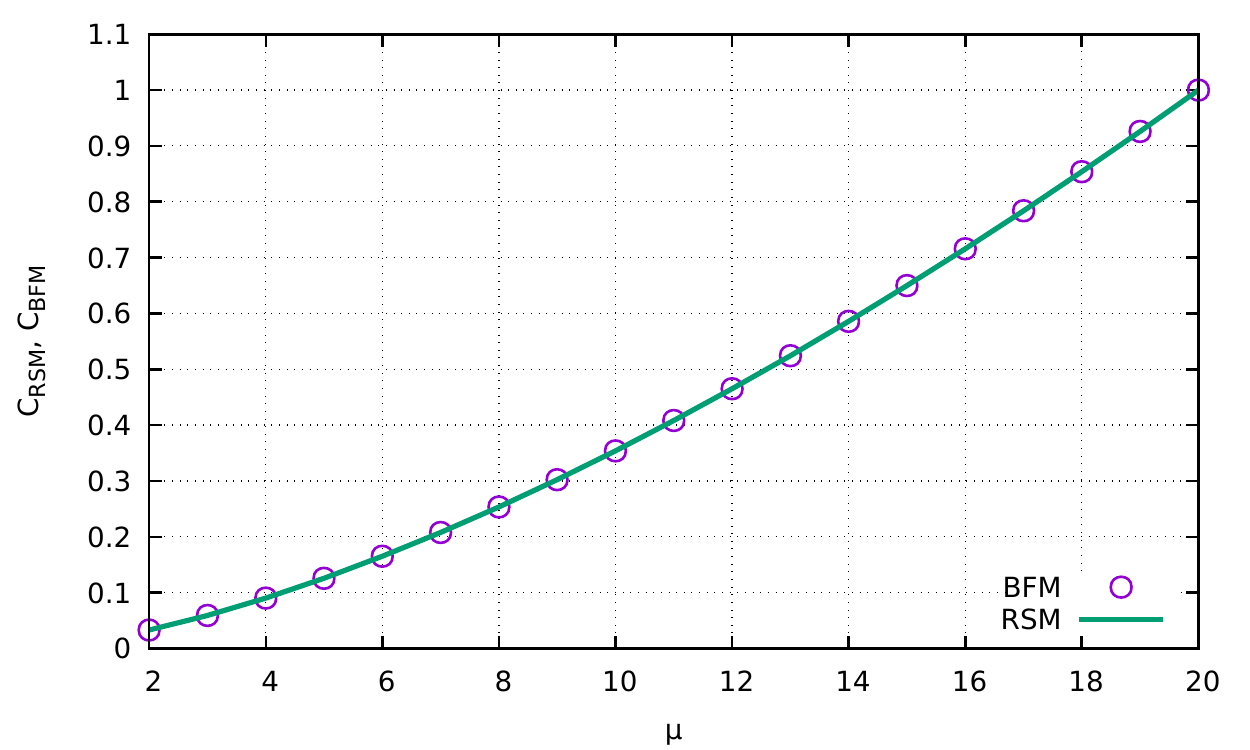}
  \includegraphics[width=\columnwidth]{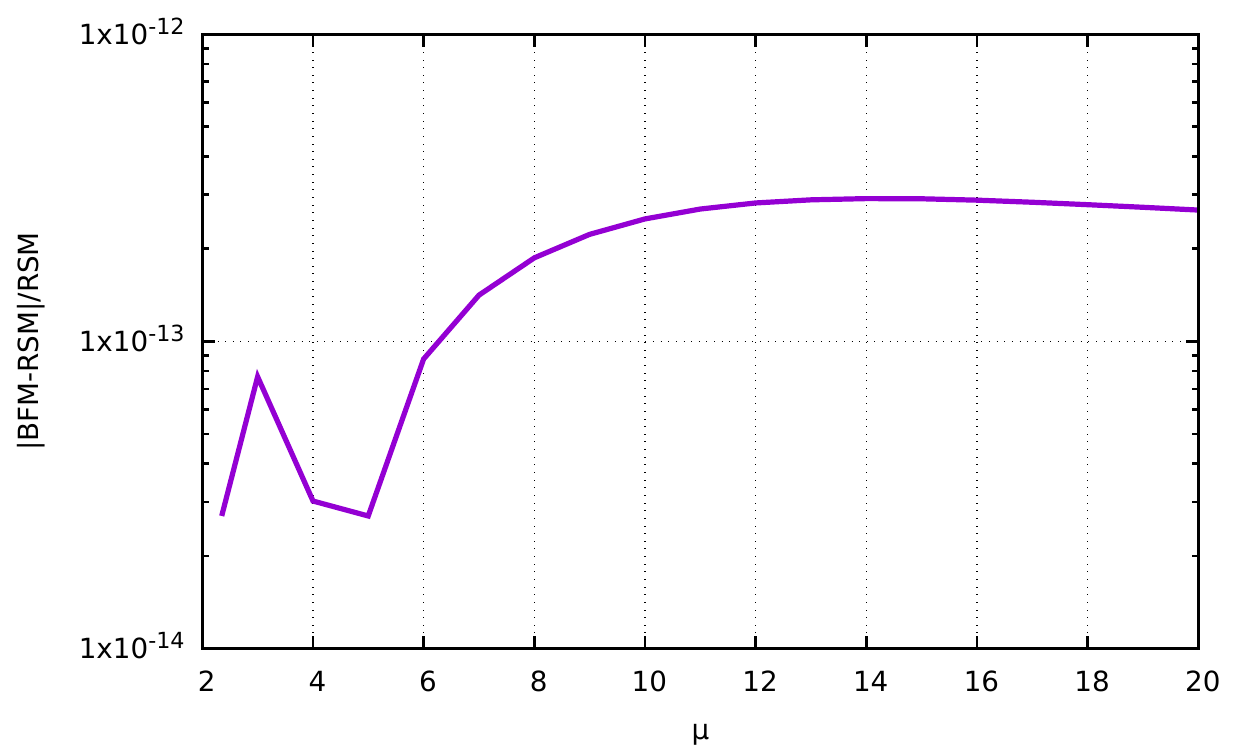}
  \caption{Solution of the $C(\mu)$ equation using the basis function method as compared with a solution computed by stepping through the $\mu$ values using a recursive approach. The upper panel shows both solutions plotted together. The lower panel depicts the absolute difference between the two solutions.}
  \label{fig:fd1}
\end{figure}
\end{center}

\subsection{Full 2D Solution: Evolution with Basis Functions}\label{2DBFM}
The numerical solution computation for the full 2D non-separable case, takes a similar approach. The solution is again taken to be a weighted-sum of the basis functions Eqn.~\ref{basis}, but with $\tau$ dependent weights. This leads to Eqn.~\ref{CS} taking the form 
\begin{eqnarray}\label{CS-mod}
&& \nonumber \sum^{N}_{n=0}\omega_n(\tau-2\dc)(\Phi_n(\mu+2\db)-\Phi_n(\mu-2\db)) = \\
&& \nonumber \hskip-0.2cm\tfrac{\sqrt{|\tau|} + \sqrt{|\tau + 2 \dc|}}{\sqrt{|\tau|} + \sqrt{|\tau - 2 \dc|}} \left(\Psi_{\mu + 2 \db, \tau + 2 \dc} - 
\Psi_{\mu - 2 \db, \tau + 2 \dc}   \right) \\
&& \nonumber \hskip-0.2cm + \tfrac{1}{2\db} \, \tfrac{\sqrt{|\tau+\dc|} - \sqrt{|\tau - \dc|}}{\sqrt{|\tau|} + \sqrt{|\tau - 2 \dc|}} \bigg[ (\mu+2\db)\Psi_{\mu + 4 \db, \tau} + \\
&& \hskip0.4cm (\mu-2\db)\Psi_{\mu - 4 \db, \tau}  - 2 \mu (1 + 2 \gamma^2 \db^2)  \Psi_{\mu,\tau} \bigg] ~.
\end{eqnarray}
The solution can then be reconstructed on the full range of $\mu, \tau$ values via
\begin{equation}\label{interp}
	\Psi_{\mu,\tau} = \sum^N_{n=0} \omega_n(\tau) \Phi_n(\mu) ~.
\end{equation}
The $\tau$ dependence is captured by stepping through a $\tau$-valued grid of this reconstructed solution.

\begin{figure}[htb!]
    \includegraphics[width=\columnwidth]{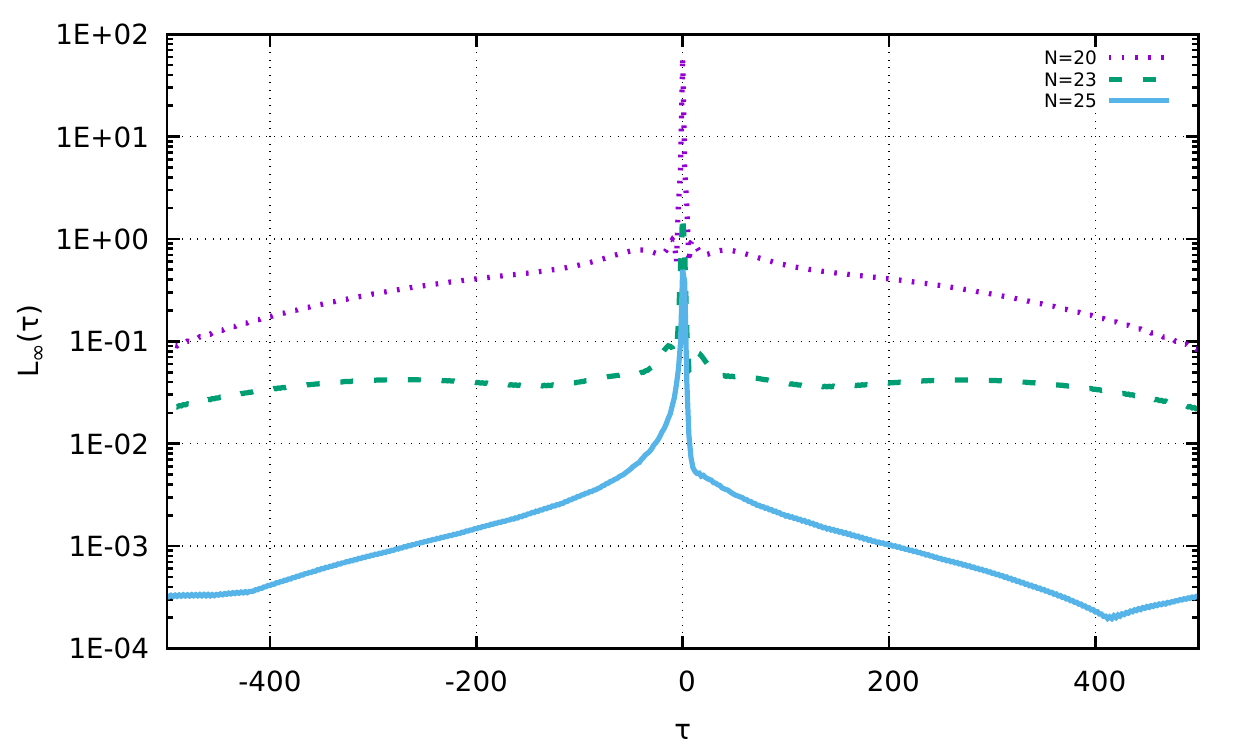}
	\caption{$L_\infty$-norm of the error at each value of $\tau$ for different number of basis elements $N=20,23,25$.}
	\label{error}
\end{figure}

Similar to the 1D variable-separated case, the basis function method ultimately leads to a linear system of equations. The system involves 
finding the  weights $\omega_n$, while the invertible system matrix represents the left-hand-side of Eqn.~\ref{CS-mod}. A boundary 
condition is implemented in the manner of row-column injection used previously, where $\lim_{\mu\rightarrow\infty}\Psi_{\mu,\tau}=0$ is represented 
as $\sum_n\omega_n\Phi_n(\mu=M)=0$. As typically done in such models, we impose a constraint that represents an initial ``semi-classical'' wave-packet, 
i.e. a Gaussian profile for the solution at large $\tau$, and then step backwards in $\tau$, ultimately evolving the system deep into the quantum regime 
and beyond.
\begin{center}
\begin{figure}[htb!]
  \includegraphics[width=\columnwidth]{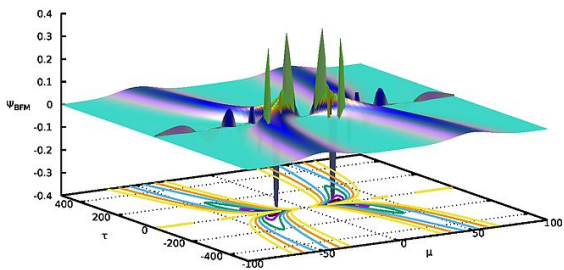}
  \includegraphics[width=\columnwidth]{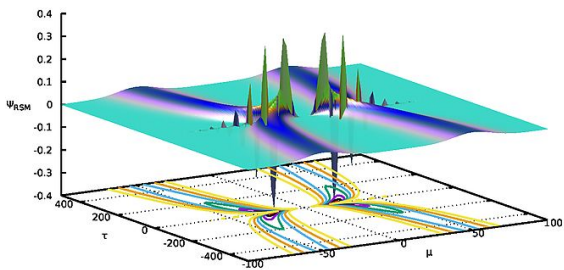}
  \caption{Solution of Eqn.~\ref{CS} using the basis function method (upper panel) as compared with a solution computed by stepping 
  through the $\tau, \mu$ values using a recursive approach (lower panel).}
  \label{fig:fd2}
\end{figure}
\end{center}
The error denoted by the $L_\infty$-norm as the maximum value of the residual between the recursive step method and basis method 
solutions at slices of constant $\tau$ is depicted in Fig.~\ref{error}.
It is clear that the error stays low for the larger values of $\tau$, however it increases in the neighborhood of $\tau=0$. This is likely 
due to the use of a relatively small number of basis elements. In fact, as seen in Fig.~\ref{error}, it is clear that overall error reduces 
dramatically with even a modest increase in the number of basis elements. Of course, one can also envision a minor tweak in the form of the
basis to reduce the error in the small $\tau$ regime further. We do not attempt to do that in this work.

\begin{center}
\begin{figure}[htb!]
  \includegraphics[width=\columnwidth]{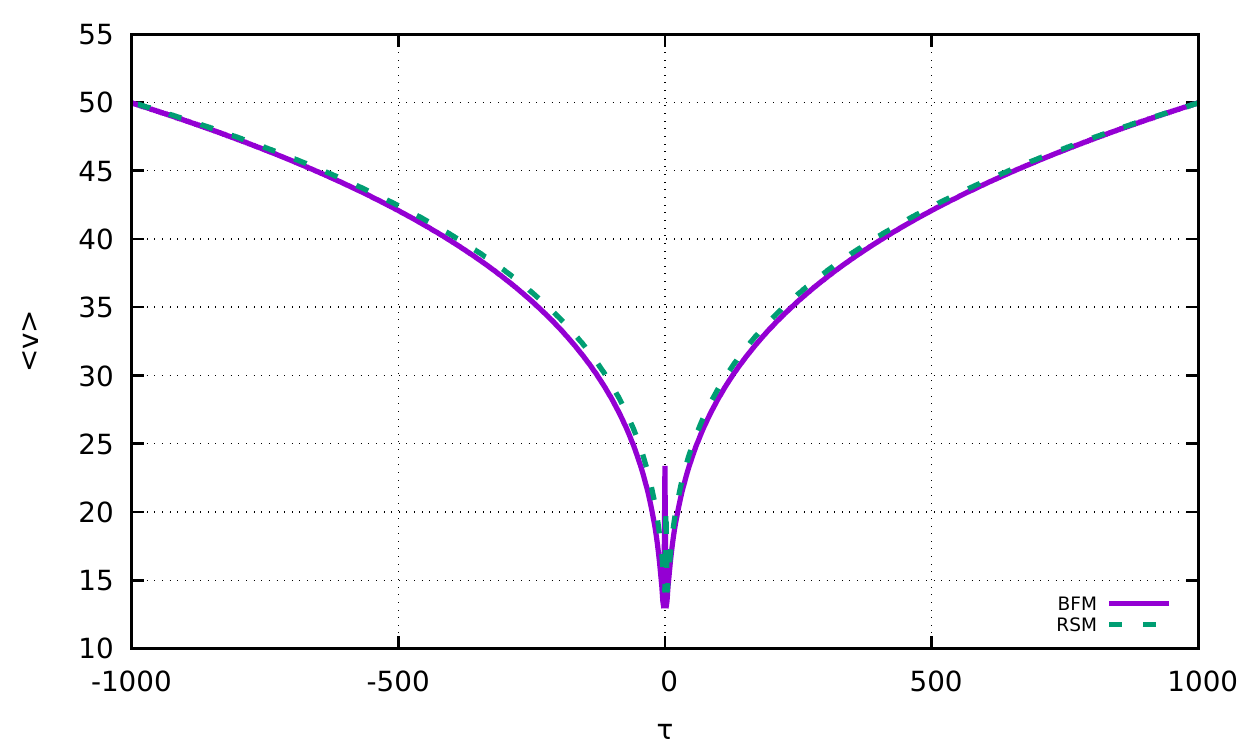}
  \includegraphics[width=\columnwidth]{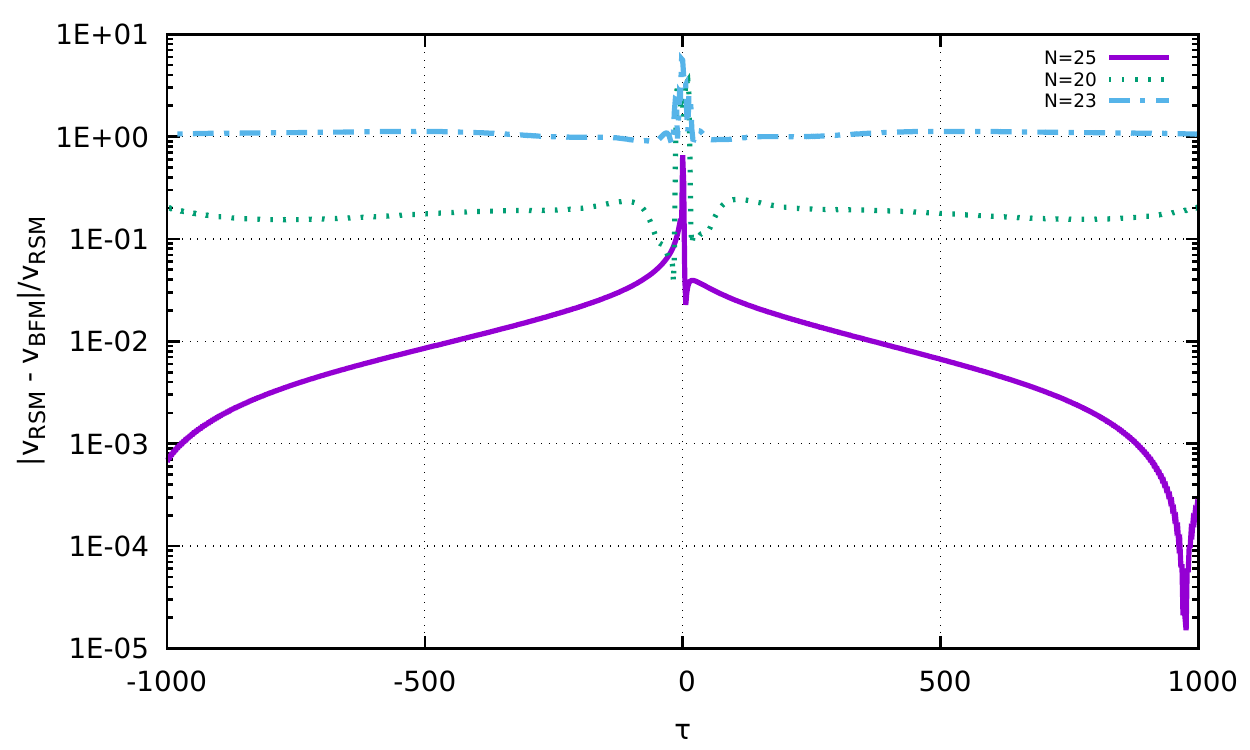}
  \caption{$\langle v\rangle$ computed using both the basis function method and the recursive step method, and the associated relative difference.}
  \label{fig:expec}
\end{figure}
\end{center}

The agreement between the basis and recursive method can be even further inspected by examining the volume expectation value $\langle v\rangle$.
This is shown in Fig.~\ref{fig:expec}.

\subsubsection{Optimization}\label{OPT}
\begin{center}
\begin{figure}[htb!]
	\includegraphics[width=\columnwidth]{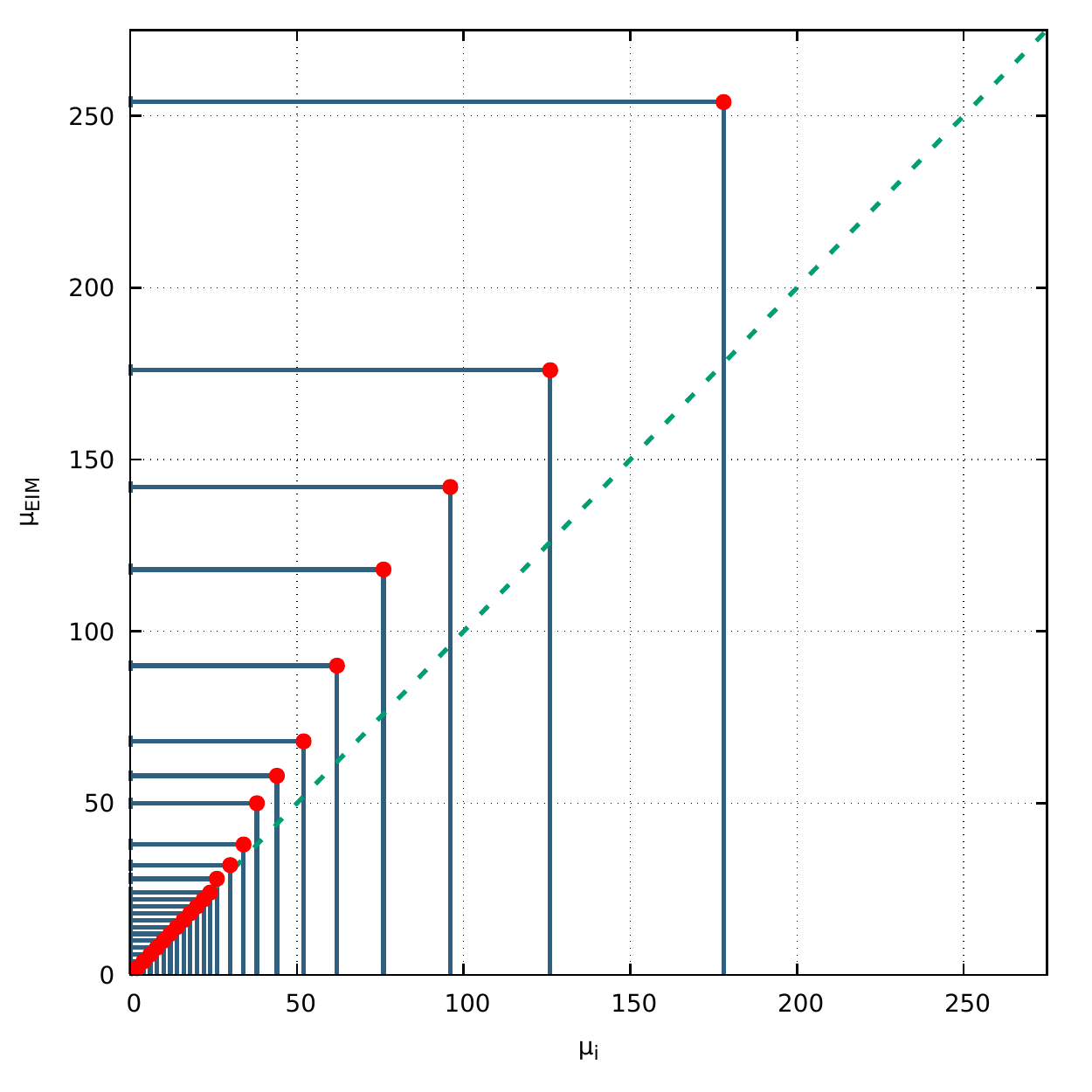}
	\caption{Subset of $\mu$ variable grid values compared with optimal $\mu$ variable grid values obtained through the empirical interpolation method.}
	\label{fig:nodes}
\end{figure}
\end{center}

To reduce computational cost, we make use of a key property of the solutions, i.e. they {\em must} get smoother for the larger triad values. 
When the solutions are smooth, they can be interpolated very effectively, thus allowing for a significant reduction in the ``sampling rate''. In 
order to take advantage of this, we only solve over a subset of triad values as computed through this expression
\begin{equation}\label{nodes}
	\mu_{i+1} = \mu_i + \lfloor 1 + \big(\frac{2\mu_i}{25}\big)^2\rfloor 
\end{equation}
where $\lfloor \rfloor$ denotes the ``integer part'' or the {\em gint} function. Visually, this set can be represented as shown in Fig.~\ref{fig:nodes}. 
Through some experimentation, we discovered that this form yields significant computational benefit with relatively little loss of accuracy. This  
allows us to reduce the effective grid size by a factor of 10, offering us a tremendous speed-up!  

In addition, while the choice of basis etc. was made largely based on physical considerations and numerical experimentation, we can show that this choice is reasonably optimal using a {\em reduced basis} approach~\cite{BasReduc2010} with an {\em empirical interpolation method} framework~\cite{EIM2008}, calculated via the {\em greedy algorithm}~\cite{AFHNT13}. This was done using the open source code {\em rompy}~\cite{rompy1,rompy2} which demonstrated that the optimal sampling nodes have a similar distribution to the nodes generated with Eqn.~\ref{nodes}. This is shown graphically in Fig.~\ref{fig:nodes}. Utilizing the chosen basis function in Eqn.~\ref{basis}, one can generate an associated Vandermonde matrix, with the elements
\begin{equation}
	V_{i,j} = \Phi_{j}(\mu_i) ~.
\end{equation}
If the basis elements $\Phi_j (\mu_i)$ are considered as a reduced basis, one can then perform a singular value decomposition (SVD) of $V_{i,j}$.
This demonstrates that the chosen basis is approximately orthonormal and well-conditioned. This was further verified by performing Gram-Schmidt 
orthonormalization on the basis, which led to no perceived difference in the orthogonality of the basis, and only an insignificant reduction in the conditioning. 
Thus, we observe that while certain techniques may be used to improve our basis function method (e.g. Gram-Schmidt, Empirical Interpolation), they 
are not necessarily required.

\subsection{Instability}\label{INST}
After performing von-Neumann stability analysis on Eqn.~\ref{CS}, it was noted previously~\cite{YKS} that the model is subject to an instability condition, 
\begin{eqnarray}
 \mu & > & 4\tau ~.
\end{eqnarray}
This condition $\mu > 4\tau$ is of particular interest as it only appears through use of von-Neumann stability analysis in the context of the full 
2D system Eqn.~\ref{CS}, and not in the variable-separable case. This is indicative of the fact that the $\mu > 4\tau$ 
condition is a consequence of the full 2D form of the quantum Hamiltonian constraint. In this section, we attempt to understand how exactly this 
instability appears in the 2D equation -- both, in the context of the stencil-based, finite-difference-like stepping approach and, of course, the 
basis function method. 

\subsubsection{The Large $\mu, \tau$ Limit}\label{CFLARG}
Beginning with the original Eqn.~\ref{CS} and taking the approximation $(\mu,\tau) >> (\db, \dc)$ and setting $\gamma\db \rightarrow 0$, we obtain 
\begin{eqnarray}\label{pdecnvrt}
&& \nonumber (1 + \sqrt{|1+ 2 \frac{\dc}{\tau}|}) \left(\Psi_{\mu + 2 \db, \tau + 2 \dc} - 
\Psi_{\mu - 2 \db, \tau + 2 \dc} \right) \nonumber \\
&& \nonumber \hskip-0.2cm + \tfrac{1}{2\db} \, (\sqrt{|1+ \frac{\dc}{\tau}|} - \sqrt{|1 -  \frac{\dc}{\tau}|}) \bigg[(\mu)\Psi_{\mu + 4 \db, \tau} + \nonumber\\
&& \nonumber \hskip0.4cm(\mu)\Psi_{\mu - 4 \db, \tau}  - 2 \mu \Psi_{\mu,\tau} \bigg]+ \nonumber\\
&& \nonumber \hskip-0.2cm (1 + \sqrt{|1 - 2 \frac{\dc|}{\tau}}) \left(\Psi_{\mu - 2 \db, \tau - 2 \dc} - 
\Psi_{\mu + 2 \db, \tau - 2 \dc} \right) \\ && = 0 ~.
\end{eqnarray}
Next, we insert the following Taylor approximation, making implicit use of the expectation that the solution tends towards smooth behavior for 
large values of $\mu, \tau$ 
\begin{align}
	\Psi_{\mu\pm l\db,\tau \pm m\dc} \approx& \Psi(\mu\pm l\db,\tau\pm m\dc)\nonumber \\
	\rightarrow& \Psi \pm l\db\tfrac{\partial}{\partial \mu}\Psi \pm m\dc\tfrac{\partial}{\partial \tau} \Psi \nonumber\\ 
	&+\tfrac{(l\db)^2\partial^2}{2!\partial \mu^2} \Psi + \tfrac{(m\dc)^2\partial^2}{2!\partial \tau^2} \Psi \nonumber\\ 
	& \pm 2lm\db\dc\tfrac{\partial^2}{\partial \mu\partial\tau}\Psi + \dots \nonumber\\
	\sqrt{|1 \pm 2\frac{\dc}{\tau}|} \approx& 1 \pm \frac{\dc}{\tau} ~. \nonumber
\end{align}
This substitution reduces Eqn.~\ref{pdecnvrt}, to the expression
\begin{align}
	4\tau \frac{\partial^2}{\partial\mu\partial\tau}\Psi + \frac{\partial}{\partial \mu}\Psi + \mu \frac{\partial^2}{\partial \mu^2} \Psi = 0
\end{align}
which can be further reduced into an equation of the form of the {\em advection} partial differential equation, 
\begin{align}\label{CFL}
	\dot\beta + \frac{\mu}{4\tau}\beta' = 0\nonumber\\
	\beta \equiv \mu \frac{\partial}{\partial\mu}\Psi ~.
\end{align}
Now, it is relatively easy to see why an instability appears for $\mu > 4\tau$ part of the computational domain. Making an analogy with the so-called, ``CFL condition''~\cite{cfl} that commonly appears in numerical partial differential equation methods, Eqn.~\ref{CFL} presents an advection equation with a locally defined speed of $v=\tfrac{\mu}{4\tau}$. This defines, at each point in ($\tau, \mu$), a local light cone extending into the past defined by the region traced out by the characteristic speed $v$. However, the finite-difference stencil as defined by the Hamiltonian constraint, has a triad stepping 
ratio of $\db / 2\dc = 1$. This results in an instability when $v>1$, since the local physical domain of dependence is no longer fully contained within the local computational domain as determined by 
the quantum Hamiltonian constraint. 

\begin{center}
\begin{figure}[htb!]
  \includegraphics[width=\columnwidth]{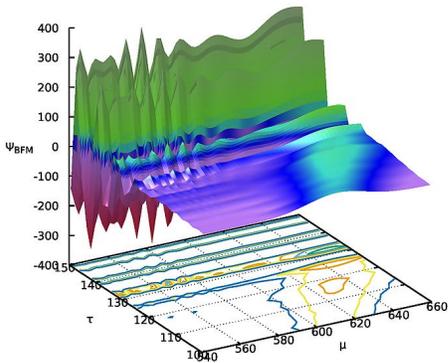}
  \caption{Evolution using the basis function method into the unstable regime of $\mu,\tau$.}
  \label{fig:boost}
\end{figure}
\end{center}

As mentioned before, the instability condition of $\mu>4\tau$ is a result of recursive stepping through both $\mu$ and $\tau$ as in Eqn.~\ref{CS}. 
With a global method like the basis method, it could be reasoned that such a local instability condition may be avoided. However, it does not; this can be seen by performing a ``shift'' of coordinates, $\mu \rightarrow \mu + 555, \tau \rightarrow \tau + 100$ and allowing part of the Gaussian wavepacket to 
evolve into the unstable regime. Fig.~\ref{fig:boost} depicts such a sample evolution. 

We find that the instability condition as shown is still applicable even in the basis function method. It is suspected that this is due to an 
intrinsic issue with the CS model, or perhaps due to the lack of a complete Hilbert space of solutions or possibly even a reflection of some underlying 
physics. Thus, von-Neumann analysis continues to be beneficial towards providing insight on the viability of models in loop quantum cosmology. More specifically, it is able to provide various constraints on such models based on the requirement that evolutions must be stable~\cite{nlqc,YKS,SS}.

\subsubsection{Stability in the Basis Function Method}\label{BFMINSTAB}
 To understand the instability in the context of the basis function approach, we take Eqns.~\ref{CS} and \ref{interp} together, while noting that 
 the only time-varying component in Eqn.~\ref{interp} is $\vec{\omega}$. We then rewrite Eqn.~\ref{CS} using vector and matrix notation
 \begin{align}
     \Psi_{\mu\pm k\db,\tau\pm n\dc} \rightarrow& \sum^N_{i=0}\Phi_i(\mu\pm k\db)\omega_i(\tau\pm n\dc)\nonumber \\
     =& \mathbf\Phi(\mu\pm k\db)\vec\omega_{\tau\pm n\dc}~.
 \end{align}
where $\mathbf\Phi$ is a square matrix with the different basis elements arranged in columns and each row representing a different $\mu$ value. 
In this notation the CS model can be rewritten as 
\begin{widetext}
\begin{eqnarray}
&& \nonumber \big( \tfrac{\sqrt{|\tau|}+\sqrt{|\tau-2\dc|}}{\sqrt{|\tau|}+\sqrt{|\tau+2\dc|}}\big)\bigg[ \mathbf\Phi(\mu-2\db)-\mathbf\Phi(\mu+2\db)\bigg]\vec\omega_{\tau-2\dc}\\
&& \nonumber \hskip-.2cm+\frac{1}{2}\big( \tfrac{\sqrt{|\tau+\dc|}-\sqrt{|\tau-\dc|}}{\sqrt{|\tau|}+\sqrt{|\tau+2\dc|}}\big)\bigg[(\mu+2\db) \mathbf\Phi(\mu+4\db)+(\mu-2\db) \mathbf\Phi(\mu-4\db)-2\mu \mathbf\Phi(\mu) \bigg]\vec\omega_{\tau}\\
&& \hskip-.2cm - \bigg[\mathbf\Phi(\mu-2\db)- \mathbf\Phi(\mu+2\db)\bigg]\vec\omega_{\tau+2\dc}=\vec 0~.
\end{eqnarray}
Given that $\mathbf\Phi$ is a matrix of full rank, we can compute the inverse of matrix $\mathbf A$ (defined below) and obtain 
\begin{align}\label{rbfgrowth}
\nonumber	\mathbf I\vec\omega_{\tau+2\dc} + \frac{1}{2}\big( \tfrac{\sqrt{|\tau-\dc|}-\sqrt{|\tau+\dc|}}{\sqrt{|\tau|}+\sqrt{|\tau+2\dc|}}\big) \mathbf{A}^{-1}\mathbf B\vec\omega_\tau-\big( \tfrac{\sqrt{|\tau|}+\sqrt{|\tau-2\dc|}}{\sqrt{|\tau|}+\sqrt{|\tau+2\dc|}}\big)\mathbf I\vec\omega_{\tau-2\dc} =&\; \vec0 \\
\nonumber	\mathbf\Phi(\mu-2\db)-\mathbf\Phi(\mu+2\db)\equiv&\;\mathbf A\\
	\big[(\mu+2\db)\mathbf\Phi(\mu+4\db) +(\mu-2\db)\mathbf\Phi(\mu-4\db) -2\mu\mathbf\Phi(\mu) \big]\equiv&\;\mathbf B ~.
\end{align}
\end{widetext}
This can be further reduced down to 
\begin{align}\label{rbfroot}
\nonumber	\mathbf I\vec\omega_{\tau+2\dc} + \frac{1}{2}\alpha \mathbf D\vec\omega_{\tau} - \beta\mathbf I \vec\omega_{\tau - 2\dc} =& \vec 0 \\
	\nonumber \mathbf D \equiv\; \mathbf A^{-1}\mathbf B&  \\
	\nonumber \alpha \equiv \big(\tfrac{\sqrt{|\tau-\dc|}-\sqrt{|\tau+\dc|}}{\sqrt{|\tau|}+\sqrt{|\tau+2\dc|}} \big)& \\
    \beta \equiv \big(\tfrac{\sqrt{|\tau|}+\sqrt{|\tau-2\dc|}}{\sqrt{|\tau|}+\sqrt{|\tau+2\dc|}} \big)&~.
\end{align}
Finally, if the following definitions are made,
\begin{align}\label{rbfreduc}
    &\vec q_{\tau} \equiv 
    \begin{bmatrix}
    \vec \omega_{\tau} \\
    \vec \omega_{\tau-2\dc}
    \end{bmatrix} &
    \mathbf \Pi \equiv 
    \begin{bmatrix}
     -\tfrac{1}{2}\alpha\mathbf{D}  & \beta\mathbf I  \\
      \mathbf I & \mathbf 0 
    \end{bmatrix}&
\end{align}
then one can use Eqn.~\ref{rbfreduc} to express Eqn.~\ref{rbfroot} in the following manner,
\begin{align}\label{rbfstab}
    \vec q_{\tau + 2\dc} = \mathbf \Pi \vec q_{\tau}~.
\end{align}
Now, we use the eigenvalues of matrix $\mathbf \Pi$ to test for stability; if any of the eigenvalue magnitudes exceed unity, that makes  
the solution of Eqn.~\ref{rbfstab} grow unboundedly. With some additional simplifications (in particular, $(\mu,\tau) >> (\db, \dc)$) 
accompanied with some numerical computations it is not difficult to see that one obtains the same unstable region ($\mu > 4\tau$) as 
derived before in Ref.~\cite{YKS} using basic von Neumann stability analysis. 


\section{Discussion and Conclusions}\label{DnC}
In this paper, we have developed a new numerical method that is particularly suitable for computing solutions to the quantum Hamiltonian constraint 
of models in loop quantum cosmology. The method is inspired by the spectral collocation method that is commonly used for numerically solving 
partial differential equations. It uses a special set of basis functions that take full advantage of the expected behavior of physical solutions 
of the model. We also compared the new basis function method with the previously used approach of a stencil-style computation using a recursive 
computation over a grid of allowed values for the triad variables. In addition, we also discussed how the stability analysis appears in the 
context of the basis approach. 

In this section we will document how the basis function method improves upon previous approaches. As pointed out earlier, the main advantage is 
computational efficiency and ease of parallelization. 

\subsection{Parallelizability}\label{PARA}
The basis function method ultimately involves the computation of a solution of a linear system of equations of size proportional to the square of the number of basis elements in use. Since this is a very well-studied problem, there exist many highly optimized solvers available even for parallel hardware 
such as multi-core CPUs and many-core GPUs. On the other hand, the recursive step method is intrinsically serial and rather challenging to parallelize. 
\begin{table}[htb!]
    \centering
    \begin{tabular}{c|c}
      Cores & BFM Run-Time (s)\\
      \hline
    1 & 161.110\\
    2 & 85.805\\
    4 & 48.852\\
    8 & 29.751\\
    \end{tabular}
    \caption{Run-Time by number of cores with simple OpenMP multi-threading on $\mu,\tau\in [-1600,1600]$ grid}
    \label{tab:performance}
\end{table}

With a simple multi-threaded implementation for the BFM algorithm, the benefit of parallelization can be clearly seen. Increased benefit is expected when the number of basis elements gets larger; and we anticipate a more in-depth study of this feature for both multi-core CPUs and many-core GPUs in future work.

\subsection{Precision}\label{PREC}
As mentioned above, the basis function method involves solving a linear system of equations. Performing an analysis on the backward error of matrix inversion techniques yields
\begin{align}
	\mathbf{A}\vec{x} =& \vec{b} \nonumber \\
	\epsilon_{BFM} = ||\mathbf{A}^{-1}\vec{b} - \vec{x}|| =& \mathcal{O}(\kappa_2\epsilon_{machine}) \nonumber \\
	\kappa_2 \equiv& ||\mathbf{A}^{-1}||_2||\mathbf{A}||_2 \propto N ~.
\end{align}
On the other hand it has been found that for a recursive step method with $\mu \in [\tfrac{-M}{2},\tfrac{M}{2}]$ and $\tau \in [\tfrac{-T}{2},\tfrac{T}{2}]$, the error $\epsilon_{RSM}\propto \mathcal{O}(T \times M\epsilon_{machine})$. Since standard backward error is 
proportional to the number of basis elements, we have shown that a small and constant number can compute solutions on a range of domains 
accurately; it stands to reason that as long as the non-zero regions of the computed solution are reasonably well sampled by the basis, the BFM 
approach can calculate a larger range of solutions with a set level of computational precision. 

\begin{widetext}
\begin{center}
\begin{figure}[htb!]
  \includegraphics[width=.45\columnwidth]{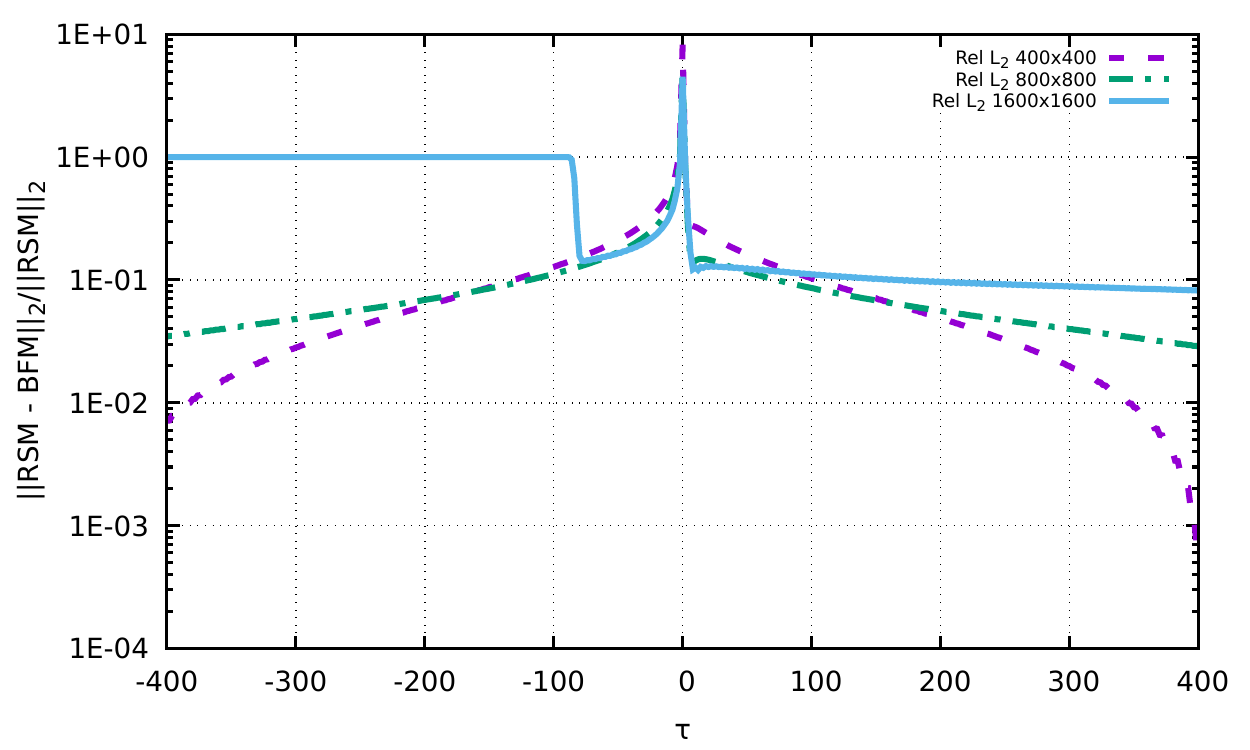}
  \includegraphics[width=.45\columnwidth]{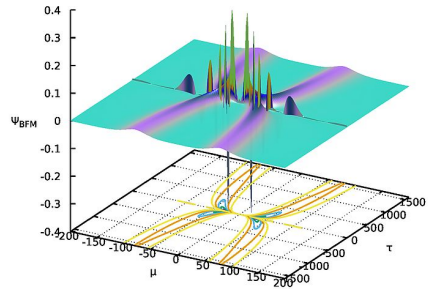}
  \caption{Relative $L_2$ error for various size domains, and evolution with the basis function method over large domain, $\mu,\tau \in [-1600:1600]$}
  \label{fig:rbflong}
\end{figure}
\end{center}
\end{widetext}

\begin{center}
\begin{figure}[htb!]
  \includegraphics[width=\columnwidth]{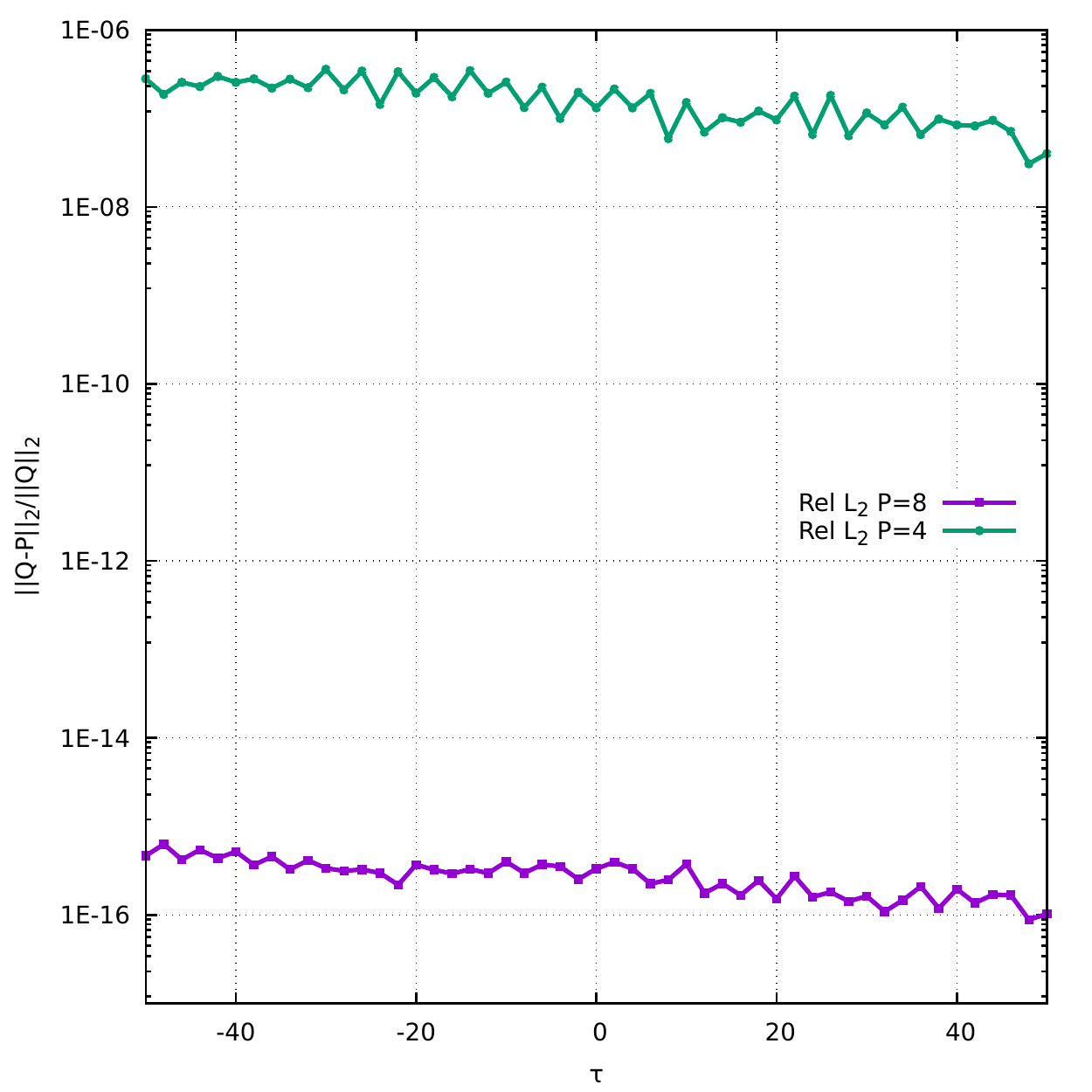}
  \caption{Precision test for single (4-byte), and double (8-byte) precision. It is clear that error grows significantly as the evolution progresses (recall, the system evolves backward in $\tau$).}
  \label{fig:prectest}
\end{figure}
\end{center}

It can be seen in Fig.~\ref{fig:rbflong} that the recursive step method based solution experiences unbounded growth due to fixed, finite 
available precision; while the basis method computed solution maintains the correct overall behavior expected. For a closer analysis of the 
effects that finite precision can have on the recursive step method solutions a precision test can be performed. For such a test, we maintain  
a constant domain and solve {\em identical} simulations with varying precision. The error can then be analyzed by taking the highest precision 
solution as {\em most accurate}. Our standard error analysis follows as 
\begin{align*}
	\epsilon_P(\tau_i) &= \frac{\sum_n |\Psi_{Quad}(\mu_n,\tau_i)-\Psi_{P}(\mu_n,\tau_i)|^2}{\sum_n \Psi^2_{Quad}(\mu_n,\tau_i)} 
\end{align*}
where $P$ is a precision other than quadruple-precision (e.g. double, single). The results are depicted in Fig.~\ref{fig:prectest}. They 
suggest that the recursive step method is far more prone to suffer from precision limitations over the basis method. This ultimately translates 
into a major computational advantage in favor of the basis method.

In summary, we have developed a new numerical approach towards solving quantum Hamiltonian constraints in loop quantum cosmology. The approach 
makes use of a set of basis functions, specifically designed using key physical features of the solutions in mind. This basis method, borrows 
from the well-known spectral collocation method for solving partial differential equations. We demonstrate the efficacy of the method and compare 
it with the previously used recursive step method. We find that the basis method offers a number of benefits over the previously used approach, 
especially in the area of computational efficiency. Throughout this paper, we use the Corichi and Singh loop quantum model of the Schwarzschild 
interior for demonstration purposes.  

{\em Acknowledgements}:  The authors would like to thank Drs. Alfa Heryudono, Scott Field and Sigal Gottlieb for very helpful suggestions throughout 
this work. G. K. thanks research support from National Science Foundation (NSF) Grant No. PHY-1701284 and Office of Naval Research/Defense University 
Research Instrumentation Program (ONR/ DURIP) Grant No. N00014181255.



\end{document}